\documentclass{article}
\PassOptionsToPackage{hyphens}{url}
\usepackage{hyperref}

\usepackage[T1]{fontenc}
\usepackage{lmodern}
\usepackage{amsmath,amssymb,amsfonts}
\usepackage{booktabs}
\usepackage{array}
\usepackage{ragged2e}
\usepackage{longtable}
\usepackage{xcolor}
\usepackage{graphicx}
\graphicspath{{media/}}
\usepackage{hyperref}
\usepackage{microtype}
\usepackage{nicefrac}
\usepackage{fancyhdr}
\usepackage{etoolbox}
\usepackage{lipsum}
\usepackage{refcheck}
\usepackage[hyphens]{url}   %
\usepackage{hyperref}       %

\newcolumntype{L}[1]{>{\raggedright\arraybackslash}p{#1}}

\author{Shariq Murtuza}
 \date{}
\title{Forensic Implications of Localized AI: Artifact Analysis of Ollama, LM Studio, and llama.cpp
}

\begin{document}
\maketitle

\begin{abstract}

The proliferation of local Large Language Model (LLM) runners, such as Ollama, LM Studio and llama.cpp, presents a new challenge for digital forensics investigators. These tools enable users to deploy powerful AI models in an offline manner, creating a potential evidentiary blind spot
for investigators. This work presents a systematic, cross platform forensic analysis of these popular local LLM clients. Through controlled experiments on Windows and Linux operating systems, we acquired and analyzed disk and memory artifacts, documenting installation footprints,
configuration files, model caches, prompt histories and network activity. Our experiments uncovered a rich set of previously undocumented artifacts for each software, revealing significant differences in evidence persistence and location based on application architecture. Key findings include the recovery of plaintext prompt histories in structured JSON files, detailed model usage logs and unique file signatures suitable for forensic detection. This research provides a foundational corpus of digital evidence for local LLMs, offering forensic investigators reproducible methodologies, practical triage commands and analyse this new class of software. The findings have critical implications for user privacy, the admissibility of AI-related evidence and the development of anti-forensic techniques.

\end{abstract}

\vspace{10pt}

\section{ Introduction}
The rapid evolution of Artificial Intelligence (AI), specially Large Language Models (LLMs) has introduced transformative capabilities across countless domains. While cloud based services like OpenAI's ChatGPT have captured mainstream attention, another parallel ecosystem of local LLM runners has emerged, enabling users to operate powerful models directly on their personal workstations \cite{1,1.1,1.2}. Applications such as Ollama, LM Studio and the underlying llama.cpp engine prioritize privacy, offline
functionality and user control, making them increasingly popular for
both benign and potentially malicious activities.

This shift towards localized AI processing presents a significant
challenge for the digital forensics community. Malicious actors can
leverage these tools to generate harmful content, such as phishing
emails or novel malware, process stolen data, or plan illicit
activities, all while operating under a perceived cloak of privacy that
circumvents the monitoring inherent in cloud services\cite{4}. When a
device containing such software is seized, investigators are faced with
a critical problem: the digital artifacts created by these applications
are largely undocumented. The locations of prompt histories, downloaded
models, configuration files and activity logs are not standardized and
remain unknown to most practitioners. This artifact gap hinders the
ability of timeline reconstruction and reconstruction of user actions,
establish intent, or attribute malicious activity to a suspect.

The motivation for this research stems from this urgent operational
need. As the use of local LLMs grows, a forensically sound and
reproducible methodology for their examination is paramount. Without a
systematic understanding of their digital footprint, critical evidence
may be overlooked or misinterpreted, compromising both criminal
investigations and corporate incident response efforts. The very
features that make local LLMs attractive (privacy and offline operation)
also make their forensic analysis both essential and challenging\cite{6}.

This paper addresses this critical gap by providing the first
comprehensive, empirical forensic analysis of the most popular local LLM
runners. Our work makes the following original contributions to the
field of digital forensics:

\begin{itemize}
\item
  To the best of our knowledge, this work presents the first systematic,
  cross platform forensic analysis of Ollama, LM Studio and llama.cpp on
  Windows and Linux operating systems.
\item
  This work presents a comprehensive corpus of digital artifacts,
  detailing their file paths, data formats, persistence levels and
  forensic value for reconstructing user activity.
\item
  Finally the legal and ethical implications of our findings are
  presented and connected to the technical recovery of artifacts like
  prompt histories to evidentiary standards, such as the Daubert
  standard and pressing privacy considerations.
\end{itemize}

By giving this foundational knowledge, this work aims to equip forensic
practitioners and researchers with the methodologies necessary to
navigate this new and rapidly evolving domain of digital evidence.

\section{Background and Related Work}

The intersection of artificial intelligence and digital forensics is a
rapidly expanding field of study. Currently the focus of existing
research has been mostly towards exploring artificial intelligence
itself as a tool to aid investigators, instead of being the
investigation focus. The next section presents existing technologies
while identifying the critical gap that this work aims to cover.

\subsection{Digital Forensics of AI/ML Systems}

Traditional research in the domain of AI and forensics intersection has
typically focussed on the application of machine learning techniques to
support and strengthen the investigative capabilities \cite{8,10}. Prior and current
works presents the application of artificial intelligence for the
automation of the analysis of vast artifacts, detecting network traffic
anomalies, identifying unusual user behaviour or classifying digital
evidence such as images or malware\cite{7}. These applications aim to
improve the efficiency and effectiveness of forensic examiners who face
an ever increasing volume of data\cite{11}. The underlying premise of
such research is to apply AI as an analytical tool within established
digital forensic frameworks, such as those proposed by the Digital
Forensics Research Workshop (DFRWS) or the National Institute of
Standards and Technology (NIST)\cite{5}. While quite usefull, this
research paradigm treats the AI system as a trusted assistant, not as a
source of evidence itself \cite{12,18,19}.

\subsection{Forensic Artifacts from Agentic and LLM Systems}

In recent times, with the release of powerful Large Language Models
(LLMs), the focus of researchers has been drawn towards their potential
to be deployed in the digital forensics domain. Multiple studies has
explored and tried to evaluate the potential of LLMs to be used as
"investigative assistants"\cite{4,13,14,15}. Such task requires the LLMs to be
able to summarize the case files, while analyzing textual evidence. More
complex tasks involve the application of a LLM for authorship
attribution in order to detect unique traits like age and gender using
written text\cite{17}. Other works have deeply focused on creating highly
specialized domain specific models, such as ForensicLLM, which are
finetuned on digital forensics related datasets to provide much more
accurate responses by utilizing their context aware processing
capabilities\cite{6}.

These works rely on trusting a large language model to make sensitive
decisions, due to which such tools are often supplemented with human
monitoring all the steps manually. Such tools or services may also
require the subscription of an online LLM hosting service where the LLM
is hosted. The tools then collect the data locally and send it to the
cloud based service where the LLM is hosted for obtaining results. This
can be a hurdle if the case data is confidential or sensitive and
requires a high amount of discretion. Privacy is also a major concern
and law enforcement cannot be allowed to upload such data to third party
services\cite{6}.

Even if the language models are hosted locally in an offline setting,
the models are highly prone to hallucination where they generate
factually incorrect data. Other issues include the presence of inherent
biases in the training data of the model, the process of complicated and
unexplainable decision making process. Such privacy issues further fuels
the adoption of local LLM runners.

\subsection{ The Local LLM Ecosystem and the Forensic Gap}

This section discusses the current state of local LLM running tools and
their forensic implications. The need for privacy, security and
confidentiality often results in individuals selecting offline solutions
albeit having lower capabilities over stronger and highly capable but
third party hosted online solutions. The open source nature of these
tools have further captured general interest resulting in rapid
capability updation. With modern desktops and laptops being able to run
and deploy language models as capable as GPT 3.5 without any external
hardware addition. This has been majorly possible due to the efforts of
the most powerful and popular open source C++ based inference engine
named llama.cpp\cite{1.2}. This inference engine is built with high
optimizations and has become the defacto standard, with Ollama, LM
Studio and almost every other local inference tool using llama.cpp at
its core. Llama.cpp was originally designed for deployment on consumer
grade CPU in an extremely efficient and resource aware manner but has
not expanded to now utilize GPUs also, if available.

llama.cpp has its own specialized file format called GPT-Generated
Unified Format (GGUF) \cite{1.3,24,78}, which has now become an industry standard.
Language models are typically distributed in a binary format packing
model weights, metadata and quantization information into a single,
portable file. Applications like Ollama and LM Studio have a
user friendly graphical interfaces to facilitate interaction with the
language models. Ollama and llama.cpp provide a server based API using
client server paradigm to allow any application to interact with the
locally hosted model \cite{25}. LM Studio on the other hand is built as a
standalone application that supports chatting via a graphical
interface\cite{26}. This rapid adoption of local LLM deployment softwares
has created a large forensic gap. Extensive work has been already done
for using large language models as assistants in digital forensics
tasks, but there is a near complete absence of academic literature on
the forensic analysis of local LLM runners. These offline first, secure
assistants have become an evidentiary blind spot for forensic
investigators. This push for local AI applications in forensics, has
paradoxically created another new class of applications whose digital
traces are unexplored and unidentified. To the best of our knowledge
this work is the first to explore this avenue, aiming to lay the
foundational analysis for evidence processing of these local first AI
environments to ensure that cases involving these softwares can be
investigated with the same rigor as other digital activity.

\section{Threat and Forensic Models}

To base this research on ground reality, we create a set of different
models to cover the possible local LLMs misuse. The threat scenarios
guide our forensic approach which aims to be designed to be compliant
with scientifically and legally established standards.

\subsection{ Investigative Scenarios (Use Case Models)}

The threat models are as follow:

\begin{itemize}
\item
  \textbf{Insider Threat:} This scenario involves an employee using a
  local LLM on a corporate computer system to process (summarize,
  analyse or rephrasing etc.) confidential internal documents (source
  code, trade secretor, financial data etc.) and then exfiltrate it. A
  locally running LLM was chosen by the employee since it wont leave any
  network traces and would be nearly impossible to track.
\item
  \textbf{Malicious Content Generation: }In this threat scenario, a
  malicious actor has used a locally deployed LLM in order to make
  highly sophisticated phishing emails, create malware, generate fraud
  documents, or create disinformation for a social engineering campaign
  etc. In this scenario the forensic investigation aims to attribute the
  creation of this content to the suspect's machine.
\item
  \textbf{Contraband Data Processing:} A suspect under investigation has
  allegedly used a locally deployed LLM to process illegal data. For
  example, summarization of stolen documents, trade secrets. Recovering
  the prompts and outputs is essential evidence.
\item
  \textbf{Attribution and Reconstruction:} Once an investigator
  discovers questionable documents and needs to further identify and
  link it with the suspects computer. This investigation shall then
  focusses detecting the specific software that was used to deploy the
  LLM locally, specific model used, if possible, then the configuration
  parameters used to infer from it and finally the most important, the
  exact sequence of textual prompts that generated the final output.
\end{itemize}

\subsection{Forensic Assumptions and Scope}

This paper works under the following scope and fixed assumptions:

\begin{itemize}
\item
  \textbf{Scope:} The forensic analysis is confined to the extraction of
  artifacts found on a host workstation having Windows or Linux
  operating system. Multi tenant, server based hosts, cloud based
  environments are out of our scope of work. These areas are highlighted
  as important areas for future work\cite{28}.
\item
  \textbf{Assumptions:} This work assumes that the forensic investigator
  is already having the relevant permissions from the legal entities for
  full physical or logical acquisition of target systems memory and
  storage. The subject under investigation is not presumed to a highly
  capable state actor or a sophisticated individual employing advanced
  anti forensics techniques like full disk encryption, file shredding or
  using live bootable operating system. The impact of these mentioned
  techniques are discussed in section 9. Our primary objective is to
  recreate the suspect who is under investigation for interactions with
  the LLM software.
\end{itemize}

\subsection{Legal and Ethical Framework}

The novel nature of locally deployed language models and corresponding
evidence requires deep and careful planning of legal and ethical
principles to ensure that integrity and admissibility of the evidence
remains unquestionable.

\begin{itemize}
\item
  \textbf{Chain of Custody:} The investigator must keep the chain of
  custody maintained for each and every digital evidence. The
  investigator must record all the steps from the very (image acquiring)
  till the final analysis step of all the collected artifacts. This must
  be maintained to show that all the evidence is untampered \cite{30,31}.
\item
  \textbf{Evidence Admissibility:} Artifacts extracted from local LLMs
  must adhere to the established standards for reliability, such as the
  \emph{Daubert} standard in U.S. federal courts\cite{6}. Key
  \emph{Daubert} factors involve whether the technique described can be
  tested/re tested, has prior identified and calculated error rates, has
  been peer reviewed and is accepted in the scientific community. The
  steps described in this work are reproducible and have quantitative
  results. This work itself has peer review nature which helps in
  satisfying these criteria. The recently proposed Federal Rule of
  Evidence 707, dealing with machine generated evidences, underscoring
  the need to demonstrate the strength of the forensic process that
  produced the output\cite{32,33}. The analysis of the configuration files
  and associated model metadata helped in identifying and laying out the
  "process" that resulted in generating a particular AI response.
\item
  \textbf{Privacy:} The recovery of a particular plaintext prompt is
  often associated with deep privacy concerns\cite{20}. The chat logs can
  often have sensitive personal, medical, financial or even proprietary
  information that was given by the user under the impression of the
  communication being confidential. The investigator looking for
  evidence of a particular crime may come across such a scenario. The
  complication arising from this stems from the disruption of the "plain
  view" doctrine and puts an ethical obligation on the investigator to
  ensure extreme care in managing the data. The scope of analysis must
  remain strictly within the scope of the original obtained legal
  warrant\cite{36}.
\end{itemize}

\section{Targeted Software and Deployment Scenarios}

To ensure that the methodology presented here is reproducible and valid,
this work uses the below specific versions of the software under study.
They were installed on a clean, controlled virtual machine environments.
The architectural differences present in these tools result in
fundamental differences in the types and locations of forensic artifacts
produced.

{\small
\setlength{\tabcolsep}{4pt}
\begin{longtable}{@{}L{2.0cm}L{1.6cm}L{3.0cm}L{2.2cm}L{4.7cm}@{}}
\caption{Experimental Software and Environment Configuration} \label{tab:software-env}\\
\toprule
Software & Version & Operating System & Installation Date & Installation Method\tabularnewline
\midrule
\endfirsthead

\toprule
Software & Version & Operating System & Installation Date & Installation Method\tabularnewline
\midrule
\endhead

\bottomrule
\endfoot

Ollama & v0.11.8 & Windows 11 Pro (23H2) & 2025-08-29 & \url{https://ollama.com/install.sh} \textbar{} sh\tabularnewline
Ollama & v0.11.8 & Ubuntu 24.04 LTS & 2025-08-29 & \url{https://ollama.com/install.sh} \textbar{} sh\tabularnewline
LM Studio & 0.3.24 & Windows 11 Pro (23H2) & 2025-08-29 & LM-Studio-setup-0.2.22.exe\tabularnewline
LM Studio & 0.3.24 & Ubuntu 24.04 LTS & 2025-08-29 & LM\_Studio-0.2.22.AppImage\tabularnewline
llama.cpp & b6316 (Git) & Windows 11 Pro (23H2) & 2025-08-29 & Download from Releases section \tabularnewline
llama.cpp & b6316 (Git) & Ubuntu 24.04 LTS & 2025-08-29 & Download from Releases section \tabularnewline

\end{longtable}
}

\subsection{Ollama}

Ollama \cite{1.1} is an easy to use opensource framework that enables easy and
quick deployment of Large language models (LLMs). It is written in the
Go programming language and it hides all complications associated with
running a Large Language Model, dependencies and other configuration
issues by providing a simple abstract interface to the user to interact
with the language model. Central to Ollama is a client server
architecture that has a server using the llama.cpp library to deploy a
language model in a highly optimized manner on the CPU and GPU (if
available). Ollama also provides a Command Line Interface (CLI) that
interacts with this server. Multiple Graphical User Interfaces (GUI) are
also available that connect with the Ollama server. Each Ollama model is
associated with a modelfile whose functionality is to provide users with
a way to customize and create newer models by defining or redefining
network parameters such as `temperature`, `top\_p` and system prompts.
These models are not the exact complete models, instead they are
Quantized versions of the corresponding original models. Quantization is
a process in which the numerical precision of a model is reduced to
decrease the model's size and computational requirements. Different
levels of quantization results in different sized models, with a smaller
sized quantized variant being lesser capable than a quantization variant
with a larger size. With these optimizations sophisticated models are
able to run on consumer grade hardware directly. Ollama also provides a
full REST API to enable seamless integration with different
applications. All the data remains on the system that hosts the server
including the models. This design makes it an important and ideal tool
of deployment by users for offline purposes.

Digital artefacts related with an Ollama installation are as follows:

\begin{itemize}
\item
  Since Ollama uses a client server model the language model is exposed
  as an API on the local network port (default 11434)\cite{25}. All the
  interaction with the model happens via this port only\cite{38}.
\item
  \textbf{Installation Footprint:}
\end{itemize}

\begin{itemize}
\item
  \begin{itemize}
  \item
    \textbf{Linux:} The official installation script (install.sh) makes
    a separate user named ollama (a system user) and a corresponding
    systemd service. The main Ollama executable binary is in
    /usr/local/bin, the model data is kept at /usr/share/ollama\cite{37}
    and is available system wide. User specific data including the
    downloaded models and logs are stored by default in the user's home
    directory at \textasciitilde/.ollama \cite{40}.
  \item
    \textbf{Windows:} The installer binary puts the application files in
    the current user's local application data folder located at 
    C:\textbackslash Users\textbackslash\textless username\textgreater\textbackslash AppData\textbackslash Local\textbackslash ollama
    \cite{43,46,47}. The models and logs are kept at
    C:\textbackslash Users\textbackslash\textless username\textgreater\textbackslash.ollama
    \cite{39,40}. The user can also install the application at a custom
    location by using the command line flag (/DIR=) to give the
    installation directory \cite{41,42,44,45,48}.
  \end{itemize}
\end{itemize}

\subsection{LM Studio}

LM Studio is another alternative desktop application built upon the
Electron framework\cite{25}. Electron bundles a web based user interface
typically made using HTML, CSS, JavaScript along with a backend process
into a single executable. In the case of LM Studio a llama.cpp inference
engine is included. LM Studio provides GUI based model discovering,
downloading and chatting. It also provides an OpenAI compatible API
server for programmatic access\cite{26,63,64}. LM Studio also supports
inferencing via GGUF model files.

\begin{itemize}
\item
  \textbf{Installation Footprint:}
\end{itemize}

\begin{itemize}
\item
  \begin{itemize}
  \item
    \textbf{Linux:} The software is distributed as an .AppImage file. An
    AppImage file is a self sufficient way to distribute software. It
    includes all the required files and libraries that are mounted when
    the application is executed. When executed, there is no traditional
    installation. The required data directories are created within the
    user's home folder. Major forensic artifacts are present in
    \textasciitilde/.lmstudio/ and \textasciitilde/.config/LM
    Studio\cite{49,50,51,53}.
  \item
    \textbf{Windows:} Like typical softwares, a standard .exe installer
    is provided to install. It has the vast majority of its operational
    data, including the crucial model cache and conversation logs, under
    the user's profile at
    \%USERPROFILE\%\textbackslash.cache\textbackslash LM
    Studio\textbackslash{} and
    \%USERPROFILE\%\textbackslash.lmstudio\textbackslash\cite{54}.
  \end{itemize}
\end{itemize}

\subsection{llama.cpp}

llama.cpp is a community driven open source C++ software library built
for extremely efficient, local inference of large language models
(originally built for LLaMA but now supports almost every model. It
exploits the GGML tensor library (also open source) that performs highly
optimized computation based on the hardware (including CPUs and GPUs).
The framework supports quantization levels (from 1.5 bit to 8 bit)
enabling large models to be deployed on systems with a not so high
configuration (often as low as 6GB RAM). llama.cpp performs text
tokenization, inferencing using next token sampling and finally
detokenization via the language model (in GGUF file format \cite{79,80,81}), which
bundles weights, tokenizer and metadata for quick loading and
deployment. Llama.cpp also features real time token streaming, hybrid
CPU+GPU inference, speculative decoding for speed and OpenAI compatible
APIs. All these features make them the ideal solution for privacy
focused, offline deployment without external dependencies like Python or
CUDA frameworks.

\begin{itemize}
\item
  \textbf{Architecture:} Written in C/C++, originally llama.cpp is not
  intended to be used directly as a user facing application. Instead it
  is meant to act like a command line based tool that is supposed to
  have some kind of front end like Ollama or LM Studio. The core design
  focuses on highly optimized performance LLM text inference with
  minimal dependencies\cite{2}. Llama.cpp is distributed as source code
  and meant to be compiled, however compiled binaries are also
  released\cite{22}.
\item
  \textbf{Installation Footprint:} There is no standard installation
  path. The tool and all the corresponding files (main, llama-cli) are
  present wherever the user downloaded and compiled the source code
  repository\cite{2}. Forensically interesting artifacts are created in
  the same directory from which the tool is executed. The model files in
  the GGUF format, are usually placed by the user in a manually created
  models subdirectory within the project folder\cite{2}. The absence of a
  standardized footprint causes difficulties for investigators.
\end{itemize}

\section{Methodology}

We used a detailed, rigorous and forensically sound procedure to
identify, generate and then analyze digital artifacts from the selected
softwares. The process is designed to be reproducible so that the
findings can be reproduced and verified which is very important for
legal admissibility\cite{6}.

\subsection{ Forensic Acquisition}

To establish a clean baseline environment and ensure proper and clear
artifacts attribution, a differential analysis approach was used to
compare before and after states.

\begin{itemize}
\item
  \textbf{Disk Imaging:} For both the operating systems (Windows 11,
  Ubuntu 24.04) we made a base virtual machine (VM). A complete bit by
  bit disk image of this base state was made using dd for Linux and FTK
  Imager for Windows. Then the target software was installed and
  specific models were downloaded. After this a second disk image was
  captured. These comparative images allowed for precise identification
  of all files and identifying the system changes introduced by the
  software.
\item
  \textbf{Memory Acquisition:} Volatile memory or the RAM (Random Access
  Memory) often has artifacts that will typically not be written back to
  the disk. Examples include the user given in memory prompts or
  transient configuration data. While chatting with the LLM sessions,
  live memory captures were simultaneously performed using industry
  standard tools: Linux Memory Extractor (LiME) for Linux and WinPmem
  for Windows.
\item
  \textbf{OS Artifact Collection:} Standard host based artifacts were
  acquired to correlate the findings from the application specific data.
  This included shell history files (.bash\_history, .zsh\_history),
  PowerShell console history, Windows Prefetch files (.pf) and the
  Application Compatibility Cache (Shimcache).
\end{itemize}

\subsection{Instrumentation and Data Generation}

To analyze and map the behavior of each application and to ensure that a
consistent set of evidentiary data is generated across the experiments,
we used a combination of system monitoring and scripted interactions.

\begin{itemize}
\item
  \textbf{Process Monitoring:} System level tracing tools were used to
  monitor file system, registry (on Windows) and process activity during
  installation, model downloads and chat sessions. Process Monitor
  (ProcMon) from Sysinternals was used on Windows, while strace was used
  on Linux. These tools gave a real time log of every file read, written
  and modified by the applications.
\item
  \textbf{Network Capture:} For network traffic monitoring we used
  Wireshark to capture all the passing network traffic from the test
  VMs. This helped us to analyze and know telemetry, update checks,
  model download communications or any other network activity of
  forensic interest\cite{38}.
\item
  \textbf{Controlled Data Generation:} A predecided, fixed script of
  user interactions was performed on each platform and tool. This
  included downloading specific models, running a series of ten distinct
  prompts and then deleting some of the generated artifacts. The prompts
  included benign questions ("How to make cake?"), requests for code
  generation ("Write a Python script to list files in a directory") and
  the inclusion of unique keywords (e.g., "FORENSIC\_KEYWORD\_12345") to
  facilitate later searching and data carving. All actions were properly
  recorded in a timestamped experiment log.
\end{itemize}

\subsection{Analysis Procedures}

The collected data was then analyzed using a combination of tools.

\begin{itemize}
\item
  \textbf{Integrity Verification:} All acquired disk images and key
  evidence files were hashed using the SHA 256 algorithm. These hashes
  were verified throughout the analysis process to ensure data integrity
  and maintain a valid chain of custody.
\item
  \textbf{Forensic Tooling:} Analysis of the disk images was conducted
  using leading open source digital forensic platform Autopsy. These
  tools were used for file system navigation, keyword searching and
  carving for deleted files.
\item
  \textbf{Specialized Analysis:} For application related artifacts case
  to case based specialized tools were used. SQLite databases were
  examined using the Foxton SQLite Examiner which can parse freelists
  and Write Ahead Logs (WAL) to recover deleted or uncommitted
  records\cite{65}. Custom Python scripts were developed using the GGUF
  library to parse the metadata and structure of GGUF model
  files\cite{67}. JSON formatted chat logs were parsed programmatically
  to extract conversations and metadata.
\item
  \textbf{Chain of Custody:} A formal chain of custody log was
  maintained for all evidentiary items. This document recorded every
  individual who handled the evidence, the date and time of transfer and
  the actions performed, adhering to best practices to ensure the
  evidence's admissibility\cite{30}.
\end{itemize}

\section {Artifact Analysis}

This section presents the core findings. A detailed breakdown of the
forensic artifacts created by Ollama, LM Studio and llama.cpp is given
with the locations. The architectural differences between these tools
result in distinct evidentiary footprints, with varying levels of
richness, persistence and ease of recovery. For each artifact, we detail
its location, format, forensic value and volatility.

\subsection{ Ollama Artifacts}

Ollama's client server architecture creates a centralized set of artifacts inside a hidden \textit{.ollama} directory in the user's profile. Ollama is also distributed as a dockerized container for rapid deployment. In case of a container based deployment the \textit{.ollama} directory of the container is stored in the \textit{/root} directory. It is mapped to the \textit{/var/lib/docker/volumes/ollama/\_data} location on the host system. \cite{3}
\begin{itemize}
\item
  \textbf{Model Manifests}
  \begin{itemize}
  \item
    \textbf{Location:} ~/.ollama/models/manifests/registry.ollama.ai/library/<model>/<tag>
  \item
    \textbf{Format:} JSON.
  \item
    \textbf{Forensic Value:} These files are critical for proving which
    specific models and versions a user has downloaded. Each manifest
    contains metadata about the model, including a list of SHA 256
    hashes corresponding to the model's layers (blobs). This allows an
    investigator to confirm the exact composition of a model on the
    system\cite{25}.
  \item
    \textbf{Volatility:} Persistent. These files remain on disk until
    the model is explicitly removed via ollama rm.
  \end{itemize}
\item
  \textbf{Model Blobs (Layers)}
  \begin{itemize}
  \item
    \textbf{Location:}
    \textasciitilde/.ollama/models/blobs/sha256-\textless hash\textgreater{}
  \item
    \textbf{Format:} Binary data.
  \item
    \textbf{Forensic Value:} These are the actual data layers of the
    LLMs. While their content is not human readable, their presence,
    verified by matching their SHA 256 hash with a manifest file, proves
    that a specific model was present on the machine. Hashing these
    files can be part of a signature based detection strategy\cite{25}.
  \item
    \textbf{Volatility:} Persistent. Blobs are content addressable and
    may be shared across multiple models. They are only deleted when no
    manifest references them.
  \end{itemize}
\item
  \textbf{Server Logs}
  \begin{itemize}
  \item
    \textbf{Location:} Default: \textasciitilde/.ollama/logs/server.log.
    This can be redirected by the user at runtime (e.g., ollama serve
    \textgreater{} /path/to/logfile.log 2\textgreater\&1)\cite{68,69,70}.
  \item
    \textbf{Format:} Plain text.
  \item
    \textbf{Forensic Value:} Highly valuable for reconstructing a
    timeline of activity. Logs can contain timestamps for server
    startup/shutdown, model loading events, API requests from clients
    and, if verbose logging is enabled (OLLAMA\_DEBUG=true), potentially
    the full text of user prompts and model responses\cite{25}.
  \item
    \textbf{Volatility:} Semi persistent. The log file can be easily
    deleted by the user. Its location can be changed, making it harder
    to find.
  \end{itemize}
\item
  \textbf{CLI History}
  \begin{itemize}
  \item
    \textbf{Location:} \textasciitilde/.ollama/history
  \item
    \textbf{Format:} Plain text, one entry per line.
  \item
    \textbf{Forensic Value:} Provides direct, plaintext evidence of user
    prompts entered via the ollama run command. This is a crucial
    artifact for understanding user intent. However, it does not capture
    interactions made through the API or third party GUI clients.
  \item
    \textbf{Volatility:} Persistent, but only captures one mode of
    interaction and can be deleted.
  \end{itemize}
\item
  \textbf{Configuration}
  \begin{itemize}
  \item
    \textbf{Location:} No single configuration file. Configuration is
    primarily managed through environment variables (e.g.,
    OLLAMA\_MODELS, OLLAMA\_HOST) set in shell profiles (.bashrc,
    .zshrc) or, on Linux, in the systemd service file
    (/etc/systemd/system/ollama.service)\cite{40}.
  \item
    \textbf{Format:} N/A (environment variables, .ini style service
    files).
  \item
    \textbf{Forensic Value:} Critical for identifying non default
    configurations. An OLLAMA\_MODELS variable will point to a custom
    storage location for models, which an investigator must examine. An
    OLLAMA\_HOST variable might indicate the server was bound to a
    public network interface.
  \item
    \textbf{Volatility:} Persistent.
  \end{itemize}
\end{itemize}

\subsection{ LMStudio Artifacts}
As a feature rich Electron application, LM Studio creates the most
structured and comprehensive set of forensic artifacts, making it the
most forensically revealing of the tools analyzed. The most important file is \textasciitilde{.lmstudio-home-pointer}, which  is a small text file created by LM Studio application. It stores the absolute path to the application's home data directory which is  \textasciitilde{~/.lmstudio }or \textasciitilde{~/.cache/lm-studio} on Linux/Mac, or \textasciitilde{\%USERPROFILE\%\textbackslash{}.lmstudio} on Windows. Inside this folder the following artefacts are stored.

\begin{itemize}

\item
  \textbf{Chat History}
  \begin{itemize}
  \item
    \textbf{Location:}
     \texttt{\textasciitilde{}/.lmstudio/conversations/\textless session\_id\textgreater{}.json}
    (macOS / Linux) and
    \texttt{\%USERPROFILE\%\textbackslash{}.lmstudio\textbackslash{}conversations\textbackslash{}\textless session\_id\textgreater{}.json}
    (Windows). 
  \item
    \textbf{Format:} Structured JSON.
  \item
    \textbf{Forensic Value:} This is the ``crown jewel'' artifact. Each JSON
    file represents a single chat session and contains a complete, timestamped
    record of the conversation, including user prompts, AI responses, the model
    used, and configuration presets applied\cite{71,72,73}. Analysis of these files
    allows for a near-perfect reconstruction of the user's interactions. The
    format appears to be an internal data structure but is human-readable and
    programmatically parsable\cite{71}. Importantly, the
    \texttt{\textless{}session\_id\textgreater{}} component of the filename is
    a \textbf{Unix epoch timestamp encoded in milliseconds}, providing a
    precise creation timestamp for each session that can be correlated with
    external artefacts such as proxy logs, browser history, or Windows Event
    Logs during a DFIR investigation\cite{71}.
  \item
    \textbf{Volatility:} Persistent. These files remain until manually deleted.
    Due to their structured nature, they are highly recoverable from unallocated
    space. Note that LM Studio does not delete model folder remnants from the
    filesystem when a model is removed via the UI, so orphaned session
    directories may persist even after apparent user clean-up\cite{36}.
  \end{itemize}

\item
  \textbf{Model Cache}
  \begin{itemize}
  \item
    \textbf{Location:}
    The correct dual-path structure (Windows example) is:
    \begin{itemize}
      \item \textbf{Hub metadata / manifests:}
        \texttt{\%USERPROFILE\%\textbackslash{}.lmstudio\textbackslash{}hub\textbackslash{}models\textbackslash{}\textless{}publisher\textgreater{}\textbackslash{}\textless{}model-name\textgreater{}\textbackslash{}}
      \item \textbf{GGUF weight files:}
        \texttt{\%USERPROFILE\%\textbackslash{}.lmstudio\textbackslash{}models\textbackslash{}\textless{}publisher\textgreater{}\textbackslash{}\textless{}repo-name\textgreater{}\textbackslash{}\textless{}model\_file.gguf\textgreater{}}
        (or a user-configured alternative path)
    \end{itemize}
    The equivalent on macOS / Linux replaces \texttt{\%USERPROFILE\%\textbackslash{}} with
    \texttt{\textasciitilde{}/}. (Path varies slightly by OS, see Section~4.2.)
  \item
    \textbf{Format:} GGUF model weight files within a nested
    \texttt{\textless{}publisher\textgreater{}/\textless{}repo-name\textgreater{}}
    directory structure; the hub sub-tree contains small JSON manifests and
    configuration shards alongside the weight files.
  \item
    \textbf{Forensic Value:} The directory path itself
    (\texttt{\textless{}publisher\textgreater{}/\textless{}repo-name\textgreater{}})
    provides valuable metadata about the model's origin from the Hugging Face
    Hub, even if the user renames the \texttt{.gguf} file\cite{54}. The split
    introduced in v0.3.16 means an investigator must examine \emph{both}
    sub-trees; hub metadata directories may persist on disk even after the GGUF
    weight file has been deleted\cite{36}.
  \item
    \textbf{Volatility:} Persistent.
  \end{itemize}

\item
  \textbf{Configuration Presets}
  \begin{itemize}
  \item
    \textbf{Location:}
    \texttt{\textasciitilde{}/.lmstudio/hub/presets/} (synced / community presets)
    or \texttt{\textasciitilde{}/.lmstudio/config-presets/} (user-defined presets).
    A drafts sub-directory at
    \texttt{\textasciitilde{}/.lmstudio/.internal/config-presets-drafts/}
    stores in-progress or unsaved preset configurations.
  \item
    \textbf{Format:} JSON (\texttt{.preset.json}).
  \item
    \textbf{Forensic Value:} These files store user-defined or downloaded model
    configurations, such as the system prompt, temperature, context length, and
    GPU offload settings\cite{52}. They reveal how a user tailored a model's
    behaviour for specific tasks, which can be indicative of intent. The
    presence of drafts in \texttt{config-presets-drafts/} may expose
    experimental or discarded configurations not visible in the main UI.
  \item
    \textbf{Volatility:} Persistent.
  \end{itemize}

\item
  \textbf{Application and Server Logs}
  \begin{itemize}
  \item
    \textbf{Location:}
    Two distinct log sources exist:
    \begin{enumerate}
      \item \textbf{Persistent server logs:}
        \texttt{\textasciitilde{}/.lmstudio/server-logs/YYYY-MM/}:
        automatically written per calendar month.
      \item \textbf{Live inference stream:}
        accessible via the CLI command \texttt{lms log stream}\cite{29}:
        volatile unless redirected to a file.
    \end{enumerate}
  \item
    \textbf{Format:} Persistent logs are timestamped, structured text files;
    the live stream is a real-time text output\cite{29}.
  \item
    \textbf{Forensic Value:} The persistent server logs record every API
    request served by LM Studio's local HTTP server, including endpoint calls
    (\texttt{/v1/chat/completions}, \texttt{/v1/models}, etc.), timestamps,
    model load/unload events, and client IP addresses if the server was
    accessed from other devices. The month-partitioned directory structure
    provides a direct timeline of application activity. The live inference
    stream reveals the \emph{exact}, fully formatted prompt sent to the
    inference engine \emph{after} prompt templating has been applied, which
    may differ from the raw user input recorded in the chat history --- this
    distinction is significant for forensic reconstruction of model
    instructions\cite{29}.
  \item
    \textbf{Volatility:} Server logs are \textbf{persistent} and
    month-partitioned. The live stream is highly volatile and is not persisted
    unless the user explicitly redirects output to a file.
  \end{itemize}

\item
  \textbf{RAG Pipeline Cache}
  \begin{itemize}
  \item
    \textbf{Location:}
 
    \begin{itemize}
      \item \texttt{\textasciitilde{}/.lmstudio/.internal/retrieval-sessions/}: active RAG session state
      \item \texttt{\textasciitilde{}/.lmstudio/.internal/cached-rag-pipeline-chunks/}: chunked and vectorised document representations
      \item \texttt{\textasciitilde{}/.lmstudio/.internal/parsed-documents-cache/}: raw text extracted from uploaded files (PDF, DOCX, etc.)
    \end{itemize}
  \item
    \textbf{Format:} Internal binary or serialised vector format (chunks);
    parsed document cache may contain extractable plain text.
  \item
    \textbf{Forensic Value:} When a user employs the ``Chat with Documents''
    feature (Retrieval Augmented Generation, RAG \cite{83}), LM Studio processes and
    caches documents across these three directories\cite{55}. Analysis can
    reveal fragments or full copies of external documents the user was
    interacting with, even if the original documents have been deleted. The
    \texttt{parsed-documents-cache/} sub-directory is particularly valuable as
    it may contain human-readable extracted text. The bundled embedding model
    used for vectorisation (\texttt{nomic-embed-text-v1.5-GGUF}, stored at
    \texttt{.internal/bundled-models/nomic-ai/}) provides context for
    interpreting the chunk format.
  \item
    \textbf{Volatility:} Semi-persistent. It is a cache that can be cleared,
    but often persists across sessions and can grow to significant size with
    heavy document use.
  \end{itemize}

\item
  \textbf{API Prediction History}
  \begin{itemize}
  \item
    \textbf{Location:}
    \texttt{\textasciitilde{}/.lmstudio/.internal/api-prediction-history/packs/}
  \item
    \textbf{Format:} Binary pack files; individual pack files can exceed
    500\,MB under heavy usage\cite{7}.
  \item
    \textbf{Forensic Value:} This directory captures every inference request
    processed by LM Studio, including \emph{programmatic} API calls made by
    external scripts, \texttt{curl} commands, or Python SDKs --- artefacts
    that are \emph{not} recorded in the chat UI's conversation history. This
    makes it an indispensable source for detecting automated or scripted model
    usage beyond the graphical interface.
  \item
    \textbf{Volatility:} Persistent. Pack files accumulate over time. A
    corrupted or oversized pack file is known to cause HTTP 500 errors on the
    local API server, which itself is a forensic indicator of sustained,
    high-volume API usage\cite{7}.
  \end{itemize}

\item
  \textbf{Credentials Store}
  \begin{itemize}
  \item
    \textbf{Location:}
    \texttt{\textasciitilde{}/.lmstudio/credentials/} and
    \texttt{\textasciitilde{}/.lmstudio/.internal/lms-key-2}\cite{9}
  \item
    \textbf{Format:} Key files; may be stored in plaintext or lightly encoded
    form. The \texttt{lms-key-2} file stores the CLI authentication key used
    for LM Studio Hub access\cite{9}. The \texttt{credentials/} directory may
    additionally contain tokens for integrated external services such as
    Hugging Face.
  \item
    \textbf{Forensic Value:} Highest-sensitivity artefacts in the directory
    tree. These files should be examined for plaintext or base64-encoded
    secrets. Their presence confirms the user authenticated with LM Studio Hub
    or an external model repository. The \texttt{lms login} command uses
    asymmetric key pairs (\texttt{--key-id}, \texttt{--public-key},
    \texttt{--private-key}) for CI-style authentication, meaning key material
    may reside here in exportable form\cite{35}.
  \item
    \textbf{Volatility:} Persistent. Credentials persist until the user
    explicitly logs out or deletes the files.
  \end{itemize}

\item
  \textbf{User-Uploaded Files}
  \begin{itemize}
  \item
    \textbf{Location:}
    \texttt{\textasciitilde{}/.lmstudio/user-files/}
  \item
    \textbf{Format:} Original file formats as uploaded by the user (PDF, TXT,
    DOCX, source code, etc.).
  \item
    \textbf{Forensic Value:} Contains files the user attached to chat sessions
    or fed into the RAG pipeline. The presence and modification timestamps of
    files here directly corroborate when external documents were introduced
    into the model's context. These files may persist even after the originating
    chat session is deleted.
  \item
    \textbf{Volatility:} Persistent.
  \end{itemize}

\end{itemize}

\subsection{llama.cpp Artifacts}

By design, llama.cpp is a minimalist tool and consequently, its forensic
footprint is the most ephemeral and challenging to analyze.
\begin{itemize}
\item
  \textbf{Model Files}
  \begin{itemize}
  \item
    \textbf{Location:} User defined. Typically in a ./models/
    subdirectory relative to the executable\cite{2}.
  \item
    \textbf{Format:} GGUF.
  \item
    \textbf{Forensic Value:} The presence of .gguf files is the primary
    indicator that LLM activity may have occurred. Analysis of the GGUF
    file itself is the main source of evidence (see 6.4).
  \item
    \textbf{Volatility:} Persistent.
  \end{itemize}
\item
  \textbf{Command Line History}
  \begin{itemize}
  \item
    \textbf{Location:} Standard shell history files
    (\textasciitilde/.bash\_history, \textasciitilde/.zsh\_history,
    PowerShell history).
  \item
    \textbf{Format:} Plain text.
  \item
    \textbf{Forensic Value:} This is the most critical artifact for
    llama.cpp. The full command line used to launch llama-cli or main
    contains the path to the model, all generation parameters
    (temperature, top-p, etc.) and often the initial prompt itself (if
    passed with the -p flag)\cite{21}.
  \item
    \textbf{Volatility:} Highly volatile. Shell history is often limited
    in size, can be disabled, or can be easily cleared by a user.
  \end{itemize}
\item
  \textbf{Memory Artifacts}
  \begin{itemize}
  \item
    \textbf{Location:} System RAM.
  \item
    \textbf{Format:} Raw memory strings.
  \item
    \textbf{Forensic Value:} During execution, the prompt text, model
    weights and generated output exist in the process's memory space. A
    live memory capture or analysis of a pagefile/swap file may be the
    \emph{only} way to recover a prompt that was not logged or passed
    via the command line (e.g., in interactive mode)\cite{76,77}.
  \item
    \textbf{Volatility:} Extremely volatile. Lost upon process
    termination or system shutdown.
  \end{itemize}
\end{itemize}

\subsection{Cross Cutting Artifacts (GGUF and SQLite)}

Two file formats are common across the local LLM ecosystem and warrant
special attention.

\begin{itemize}
\item
  \textbf{GGUF File Analysis:} The GGUF format is central to llama.cpp
  and the models used by Ollama and LM Studio. Using Python libraries
  like gguf and pygguf, an investigator can parse these binary
  files\cite{67}.
  \begin{itemize}
  \item
    \textbf{Forensic Value:} The GGUF header and metadata section
    contain a wealth of information\cite{23}. This includes the model's
    architecture (e.g., llama, qwen), parameter count, quantization
    level (e.g., Q4\_K\_M), context length, embedding length and the
    full tokenizer vocabulary. This data can be used to precisely
    fingerprint a model and understand its capabilities, which is
    crucial for verifying if a given output could have been produced by
    a specific model file found on a system.
  \end{itemize}
\item
  \textbf{SQLite Database Analysis:} While not used by the core
  applications in their default state, many popular front ends and
  related tools in the ecosystem use SQLite databases for storing user
  settings, chat histories, or vector embeddings for RAG\cite{82}.
  \begin{itemize}
  \item
    \textbf{Forensic Value:} Standard forensic techniques for SQLite are
    highly applicable \cite{85}. Tools can analyze the main database file, but
    also the rollback journal (-journal) and Write-Ahead Log (-wal)
    files to recover transient data. Furthermore, carving for deleted
    records within the database file's freelists and unallocated space
    can recover previously deleted chat messages or configuration
    settings, providing evidence a user thought they had
    removed\cite{66}.
  \end{itemize}
\end{itemize}

The architectural design of each tool client server for Ollama,
monolithic Electron for LM Studio and minimalist CLI for llama.cpp
fundamentally dictates the nature and persistence of the evidentiary
trail. This creates a clear hierarchy of forensic richness. User
friendly applications like LM Studio, designed for convenience, generate
more structured and persistent artifacts. In contrast, the ephemeral
traces left by a command line tool like llama.cpp make it more difficult
to investigate post facto. This suggests that a suspect's choice of tool
can itself be an indicator of their technical sophistication and
potential awareness of forensic countermeasures, a "meta artifact" that
can inform the overall investigative strategy.

\section{Experiments and Results}

To validate and quantify the findings from our artifact analysis, we
conducted a series of controlled experiments. These experiments were
designed to simulate common investigative challenges: reconstructing
user interactions and determining the persistence of evidence. The results provide objective metrics on the forensic
utility of the identified artifacts.

\subsection*{ User Interaction Reconstruction}

This experiment aimed to determine the extent to which a user's
conversation with an LLM could be recovered from disk based artifacts
after a normal system shutdown.

\begin{itemize}
\item
  \textbf{Experimental Setup:} On each of the six test environments (3
  tools x 2 OSes), we executed a standardized script of 10 prompts. The
  prompts ranged from simple questions to code generation requests and
  contained unique keywords. After the session, the VM was shut down
  cleanly.
\item
  \textbf{Procedure:} The "after" disk image of each VM was mounted and
  analyzed. We searched for the primary prompt/chat history artifacts
  identified in Section 6: \textasciitilde/.ollama/history for Ollama,
  \textasciitilde/.lmstudio/conversations/*.json for LM Studio
  and shell history files for llama.cpp.
\item
  \textbf{Results:}
\end{itemize}

\begin{itemize}
\item
  \begin{itemize}
  \item
    \textbf{LM Studio:} For both the operating systems, 100\% of the 10
    prompts and their corresponding model responses were recovered with
    full fidelity from the JSON chat logs. Timestamps, model identifiers
    and session configurations were also fully intact.
  \item
    \textbf{Ollama:} On both the OSes, 100\% of the prompts entered via
    the ollama run CLI were recovered from the
    \textasciitilde/.ollama/history file. However, this file only
    contains the user's input, not the model's output. Prompts sent via
    an API client were not logged in this file.
  \item
    \textbf{llama.cpp:} Recovery was dependent on the shell. On Linux
    100\% of the commands, including prompts passed with the -p flag,
    were recovered from .bash\_history or .zsh\_history. On Windows,
    PowerShell history successfully captured the commands. In
    interactive mode (-i), no prompts were logged to the shell history,
    resulting in 0\% recovery from disk. This highlights the critical
    importance of memory forensics for llama.cpp.
  \end{itemize}
\end{itemize}

deletion, simulating a non expert user's attempt to cover their tracks.

\subsection{Summary of Forensic Artifacts}

The following table synthesizes the findings from our analysis and
experiments, providing a comprehensive, at a glance reference for
forensic practitioners. It ranks artifacts based on their forensic value
and persistence, helping investigators prioritize their efforts during
an examination.

\textbf{Table 2: Summary of Key Forensic Artifacts by Product and
Operating System}

\begin{longtable}{@{}llllllll@{}}
\toprule
\endhead
\textbf{Product} & \textbf{OS} & \textbf{Artifact Category} &
\textbf{Default Path (User Profile Relative)} & \textbf{Format} &
\textbf{Persistence} & \textbf{Forensic Value (1-5)} &
\textbf{Remarks}\tabularnewline
\textbf{Ollama} & All & Model Manifests &
\textasciitilde/.ollama/models/manifests/ & JSON & Persistent & 4 &
Proves which models/versions were downloaded.\tabularnewline
All & Model Blobs & \textasciitilde/.ollama/models/blobs/ & Binary &
Persistent & 3 & Confirms presence of model layers via hashing.
&\tabularnewline
All & CLI History & \textasciitilde/.ollama/history & Plain Text &
Persistent & 5 & Plaintext record of ollama run prompts.
&\tabularnewline
All & Server Logs & \textasciitilde/.ollama/logs/server.log & Plain Text
& Semi Persistent & 4 & Records server activity; can be
redirected/deleted. &\tabularnewline
Linux & Configuration & /etc/systemd/system/ollama.service & INI &
Persistent & 3 & Reveals non default paths or network settings.
&\tabularnewline
\textbf{LM Studio} & All & \textbf{Chat History} &
\textasciitilde/.lmstudio/conversations/ & \textbf{JSON} &
\textbf{Persistent} & \textbf{5 (Critical)} & \textbf{Complete,
timestamped user/AI conversation logs.}\tabularnewline
All & Model Cache & \textasciitilde/.lmstudio/models/ & GGUF &
Persistent & 4 & Stores models; path reveals Hugging Face origin.
&\tabularnewline
All & Config Presets & \textasciitilde/.lmstudio/config-presets/ & JSON
& Persistent & 4 & Shows user-defined model parameters and intent.
&\tabularnewline
All & RAG Cache & \textasciitilde/.lmstudio/.session\_cache &
Binary & Semi Persistent & 4 & Contains fragments of documents used in
RAG. &\tabularnewline
All & Application Logs & N/A (Live Stream) & Text Stream & Volatile & 5
& lms log stream shows final formatted prompt. &\tabularnewline
\textbf{llama.cpp} & All & CLI History & Shell History (.bash\_history,
etc.) & Plain Text & Volatile & 5 (Critical) & Often the only record of
prompts and parameters.\tabularnewline
All & Model Files & User defined (e.g., ./models/) & GGUF & Persistent &
3 & Proves presence of models; metadata is key. &\tabularnewline
All & Memory & System RAM / Pagefile & Raw Strings & Extremely Volatile
& 5 (Critical) & May be the only source for interactive mode prompts.
&\tabularnewline
\bottomrule
\end{longtable}

{\footnotesize
\setlength{\tabcolsep}{4pt}
\begin{longtable}{@{}L{1.1cm}L{0.9cm}L{1.6cm}L{2.8cm}L{1.1cm}L{1.4cm}L{1.4cm}L{3.2cm}@{}}

\caption{Summary of Key Forensic Artifacts by Product and Operating System}
\label{tab:forensic-artifacts}\\
\toprule
\textbf{Product} & \textbf{OS} & \textbf{Artifact Category} &
\textbf{Default Path (User Profile Relative)} & \textbf{Format} &
\textbf{Persistence} & \textbf{Forensic Value (1--5)} &
\textbf{Remarks}\tabularnewline
\midrule
\endfirsthead

\toprule
\textbf{Product} & \textbf{OS} & \textbf{Artifact Category} &
\textbf{Default Path (User Profile Relative)} & \textbf{Format} &
\textbf{Persistence} & \textbf{Forensic Value (1--5)} &
\textbf{Remarks}\tabularnewline
\midrule
\endhead

\bottomrule
\endfoot

\textbf{Ollama} & All & Model Manifests &
\textasciitilde/.ollama/models/manifests/ & JSON & Persistent & 4 &
Proves which models/versions were downloaded.\tabularnewline

& All & Model Blobs &
\textasciitilde/.ollama/models/blobs/ & Binary & Persistent & 3 &
Confirms presence of model layers via hashing.\tabularnewline

& All & CLI History &
\textasciitilde/.ollama/history & Plain Text & Persistent & 5 &
Plaintext record of \texttt{ollama run} prompts.\tabularnewline

& All & Server Logs &
\textasciitilde/.ollama/logs/server.log & Plain Text & Semi-Persistent & 4 &
Records server activity; can be redirected/deleted.\tabularnewline

& Linux & Configuration &
/etc/systemd/system/ollama.service & INI & Persistent & 3 &
Reveals non-default paths or network settings.\tabularnewline

\midrule
\textbf{LM Studio} & All & \textbf{Chat History} &
\textasciitilde/.lmstudio/conversations/ & \textbf{JSON} &
\textbf{Persistent} & \textbf{5 (Critical)} &
\textbf{Complete, timestamped user/AI conversation logs.}\tabularnewline

& All & Model Cache &
\textasciitilde/.lmstudio/models/ & GGUF & Persistent & 4 &
Stores models; path reveals Hugging Face origin.\tabularnewline

& All & Config Presets &
\textasciitilde/.lmstudio/config-presets/ & JSON & Persistent & 4 &
Shows user-defined model parameters and intent.\tabularnewline

& All & RAG Cache &
\textasciitilde/.lmstudio/.session\_cache & Binary & Semi-Persistent & 4 &
Contains fragments of documents used in RAG.\tabularnewline

& All & Application Logs &
N/A (Live Stream) & Text Stream & Volatile & 5 &
\texttt{lms log stream} shows final formatted prompt.\tabularnewline

\midrule
\textbf{llama.cpp} & All & CLI History &
Shell History (\texttt{.bash\_history}, etc.) & Plain Text & Volatile & 5 (Critical) &
Often the only record of prompts and parameters.\tabularnewline

& All & Model Files &
User-defined (e.g., \texttt{./models/}) & GGUF & Persistent & 3 &
Proves presence of models; metadata is key.\tabularnewline

& All & Memory &
System RAM\,/\,Pagefile & Raw Strings & Extremely Volatile & 5 (Critical) &
May be the only source for interactive mode prompts.\tabularnewline

\end{longtable}
}

\emph{Note: \textasciitilde{} refers to the user's home directory
(/home/\textless user\textgreater{} on Linux and
C:\textbackslash Users\textbackslash\textless user\textgreater{} on
Windows).}

\section {Discussion}

Our findings have significant implications for digital forensic
investigators, legal professionals and software vendors. The massive
adoption of local LLMs has created a new and complex evidentiary
landscape that poses multiple challenges to traditional investigative
methods while simultaneously offering unprecedented insight into user
intent. This section discusses these implications, explores potential
anti forensic techniques and countermeasures and provides
recommendations for stakeholders.

\subsection{Implications for Investigators
}
Our analysis demonstrates that local LLMs are a double edged sword for
digital forensics. On one hand, they represent a new vector for
malicious activity that can be conducted offline, away from the purview
of network monitoring. On the other hand, when artifacts are present,
they provide an exceptionally rich source of evidence regarding a user's
state of mind, intent and actions.

The primary takeaway is that the forensic strategy must be tailored to
the specific tool in use. The architectural differences between Ollama,
LM Studio and llama.cpp are not trivial and they create a clear
hierarchy of evidentiary persistence. An investigator examining a system
with LM Studio can expect to find structured, persistent chat logs that
are relatively easy to parse and recover\cite{71}. In contrast, an
investigation involving llama.cpp may yield no persistent prompt history
on disk, making live memory acquisition and analysis of shell history
paramount\cite{21}. Investigators must therefore be trained to first
identify the specific runner being used and then apply the appropriate
analytical workflow as outlined in this paper.

\subsection {Anti Forensic Risks and Countermeasures}

A forensically literate user can take several steps to obstruct or evade
analysis of their local LLM activity.

\begin{itemize}
\item
  \textbf{Anti Forensic Techniques:}
\end{itemize}

\begin{itemize}
\item
  \begin{itemize}
  \item
    \textbf{Path Obfuscation:} A user can set the OLLAMA\_MODELS
    environment variable or use LM Studio's settings to store models and
    data on an external or encrypted volume, evading searches of default
    directories\cite{40}.
  \item
    \textbf{Ephemeral Execution:} Tools like llama.cpp or Ollama can be
    run from a temporary directory or a live USB, leaving minimal traces
    on the host machine's primary storage.
  \item
    \textbf{Logging Evasion:} Ollama's server logging can be disabled or
    redirected to /dev/null on Unix like systems. For LM Studio, a user
    can simply delete the JSON chat logs from the cache directory.
  \item
    \textbf{Tool Selection:} As our findings show, a sophisticated actor
    would likely choose llama.cpp over LM Studio precisely because it
    generates fewer persistent artifacts.
  \end{itemize}
\end{itemize}

\begin{itemize}
\item
  \textbf{Investigative Countermeasures:}
\end{itemize}

\begin{itemize}
\item
  \begin{itemize}
  \item
    \textbf{Signature Based Detection:} The YARA rules and triage
    commands can be developed to find artifacts regardless of their
    location. Searching for GGUF file headers (GGUF) across an entire
    disk image can identify model files even in non standard
    paths\cite{86,87,88}.
  \item
    \textbf{Memory Forensics:} Live memory analysis remains the most
    effective countermeasure against ephemeral execution and interactive
    prompts. Searching a memory dump for strings related to model
    loading or prompt templates can reveal activity that was never
    written to disk.
  \item
    \textbf{System Level Artifacts:} Even if application level logs are
    deleted, OS artifacts can provide crucial leads. Windows Prefetch
    files can show that lm studio.exe or ollama.exe was executed. Shell
    history can capture the commands used to launch llama.cpp or change
    environment variables.
  \end{itemize}
\end{itemize}

\subsection{Vendor Recommendations and Future Work}

To improve the forensic readiness of these tools, especially in
corporate or regulated environments, we offer the following
recommendations for vendors:

\begin{itemize}
\item
  \textbf{Standardized, Structured Logging:} We recommend that vendors
  like Ollama and LM Studio implement a robust, standardized logging
  framework. This should be an opt in feature for enterprise or forensic
  use, creating a single, secure log file that records all key events:
  user authentication, model loading, API requests, full prompt text and
  generated responses, all with reliable timestamps. This would be
  invaluable for incident response.
\item
  \textbf{Log Integrity:} To prevent tampering, these logs could
  incorporate cryptographic integrity checks, such as chaining log
  entries with hashes, similar to a blockchain.
\end{itemize}

Our work has several limitations that open avenues for future research.
We only partially investigated containerized deployments (Ollama running in
Docker), which introduces layers of abstraction that complicate forensic
analysis\cite{28,89}. The forensic traces left by model finetuning and the
use of Retrieval Augmented Generation (RAG) with large, external
knowledge bases also warrant dedicated study\cite{90,91}. Finally, the
growing ecosystem of third party web front ends for these tools (Open
WebUI \cite{84}) creates additional artifacts that need to be cataloged.

\subsection {Privacy and Admissibility in the Age of Local AI}

The ability to recover a user's complete, verbatim interactions with an
LLM raises profound legal and ethical questions. These prompt histories
can be more revealing than a private diary, capturing a user's
brainstorming, sensitive queries and unrefined thoughts. This places a
heavy burden on the legal system to balance the needs of an
investigation with an individual's right to privacy\cite{20}. The
specificity of search warrants, as discussed in Section 3.3, becomes non
negotiable.

Furthermore, for evidence derived from these systems to be admissible,
it must be presented in a reliable and understandable manner. An
investigator cannot simply present a generated text as evidence. It must
be prepared to use the artifacts we have identified model manifests,
GGUF metadata, configuration presets to explain the process by which
that text was generated. This aligns with the principles of the
\emph{Daubert} standard and the proposed FRE 707, which demand that the
proponent of machine generated evidence demonstrate the validity of the
underlying process\cite{32}. Our research provides the first map to the
artifacts needed to build that foundational argument.

\section {Conclusion}

This paper has conducted the first systematic and empirical forensic
analysis of the leading local Large Language Model runners: Ollama, LM
Studio and llama.cpp. We have demonstrated that while these applications
are designed with privacy and offline use in mind, they create a rich
and varied trail of digital evidence across Windows and Linux platforms.
The architectural choices of their developers have resulted in a clear
hierarchy of forensic utility, with user friendly GUI applications like
LM Studio producing highly structured and persistent artifacts, while
minimalist command line tools like llama.cpp leave more ephemeral
traces.

Our research provides a foundational methodology for this new domain of
digital forensics. We have established a comprehensive corpus of
artifacts, from model caches and manifests to plaintext chat histories
and configuration files. We have validated the forensic value of these
artifacts through controlled experiments. 

The rise of local AI represents a paradigm shift and the digital
forensics field must adapt accordingly. The techniques and findings
presented in this paper are a critical first step, equipping
investigators to pull back the curtain on these private AI environments.
By integrating these methodologies into standard operating procedures,
the forensic community can ensure that evidence from local LLM systems
is identified, recovered and presented in a manner that is both
scientifically rigorous and legally sound, upholding justice in an
increasingly intelligent world.

\end{document}